\documentclass[%
 reprint,
 amsmath,amssymb,
 aps,
]{revtex4-2}
\usepackage{graphicx}
\usepackage{dcolumn}
\usepackage{bm}
\usepackage{braket}
\usepackage{xcolor}

\def\red#1 {\textcolor{red} {#1} }

\begin{document}


\title{Bethe Salpeter Equation Spectra for Very  Large Systems}

\author{Nadine C Bradbury}
 \affiliation{Department of Chemistry and Biochemistry, UCLA, Los Angeles CA 90095-1569 USA}

\author{Minh Nguyen}
 \affiliation{Department of Chemistry and Biochemistry, UCLA, Los Angeles CA 90095-1569 USA}
\author{Justin R Caram}
 \affiliation{Department of Chemistry and Biochemistry, UCLA, Los Angeles CA 90095-1569 USA}
\author{Daniel Neuhauser}
 \affiliation{Department of Chemistry and Biochemistry, UCLA, Los Angeles CA 90095-1569 USA}

\date{\today}

\begin{abstract}
We present a highly efficient method for the extraction of optical properties of very large molecules via the Bethe-Salpeter equation. The crutch of this approach  is the calculation of the action of the effective Coulombic interaction, $W$, through a stochastic TD Hartree propagation, which uses only 10 stochastic orbitals rather than propagating the full sea of occupied states. This leads to a scaling that is at most cubic in system size, with trivial parallelization of the calculation. We apply this new method to calculate the spectra and electronic density of the dominant excitons of a carbon-nanohoop bound fullerene system with 520 electrons, using less than 4000 core hours.
\end{abstract}

\maketitle

\section{Introduction}
\vspace{-10pt}

Accurate calculation of optical spectra for large systems is essential for future novel optical and electronic devices, in fields ranging from photovoltaics,\cite{Kippelen2009} photocatalysis,\cite{Improta2016} organic semiconductors,\cite{Coe2002,Tessler2002} to even the mechanisms of UV damage on DNA.\cite{Wu2019} The ubiquitous method of the field is time dependent density functional theory (TD-DFT), but unfortunately this method lacks the accuracy needed for predictive power in most electronically complex systems. The Bethe-Salpeter Equation (BSE) formalism is an increasingly popular alternative for calculating electronic spectra.\cite{Blase2020} The success of the BSE is due to the proper inclusion of an effective long range exchange kernel, which solves the failures of TD-DFT in accurately describe charge transfer excitations and avoided crossings.\cite{Dreuw2004,Botti2004}

Current conventional methods for solving the BSE are substantially more computationally demanding than most implementations of TD-DFT due  the explicit calculation of a large number of occupied and virtual electronic states and the evaluation of a large number of screened exchange integrals between valence and conduction states, yielding a typical scaling of $ \mathcal{O}(n_o^6)$, for $n_o$ valence orbitals.\cite{BerkeleyGW, YAMBObse, Siesta} TD-DFT with local exchange functionals has a naive scaling of $\mathcal{O}(n_o^4)$. However, progress in the field has reduced the scaling with techniques that only requires the occupied orbitals which implicitly interact with all conduction states, what we later term a ``mixed representation".\cite{Stratmann1998, Baroni2001,Baer2004,Neuhauser2005,Walker2006,Wilson2008,Ge2014-BaroniTDDFT,Zhang2015-GangTDDFT,Rocca2012-WEST} While very important for TD-DFT, the speedup gained from use of an approximate iterative method versus direct diagonalization is only worthwhile in BSE if solving for the absorption is the algorithmic bottleneck.

To go beyond TD-DFT to BSE requires constructing the effective Coulombic interaction, $W$, the most computationally expensive step. To overcome this issue, we adopt our previous stable iterative methods\cite{Baer2004,Neuhauser2005}, and pull from our previous works in both TD-DFT/BSE\cite{Baer2004,Neuhauser2005,Neuhauser2014sDFT,Rabani2015,Gao2015} and stochastic GW\cite{Neuhauser2014sGW,Vlek20181,Vlek20182,Vlek20183,StochasticGWours} to present an efficient stochastic generation of $W$ within an iterative BSE technique.  Our combined approach uses stochastic time-dependent propagation to obtain the action of $W$ on each required term in linear scaling.\cite{Neuhauser2014sGW} Overall, this results in an efficient method with at most cubic scaling with respect to system size.  The method and its application to a large organic semiconductor are detailed below.
\vspace{-10pt}

\section{\label{Method}Method}
\vspace{-5pt}
Overall, every method for solving the BSE has two numerical parts. Construction of the ``kernels", i.e., the action of the effective interaction $W$ on a given transition, and then diagonalizing or iterating the resulting Bethe-Salpeter Hamiltonian-like operator for the excitons. The full algorithm is covered here for completeness.  

The starting point is a closed shell molecular system with $2 N_{\rm occ}$ electrons.  We are interested in excitons composed of a mixture of $n_o$ occupied (valence) orbitals, $\phi_i,\phi_j,...$, times $n_c$ conduction (virtual) ones written as $\phi_a,\phi_b,...$.  Typically $n_o \ll N_{\rm occ}$ states are considered.  The orbitals are eigenstates of a zero-order Hamiltonian $H_0$, with occupied-state eigenvalues $\varepsilon_i,\varepsilon_j,...$ and virtual-state eigenvalues $\varepsilon_a,\varepsilon_b,...$.  Formally the zero-order Hamiltonian eigenvalues come from a very accurate method, specifically self-consistent GW. Additionally, we will use the well established Tamm-Dancoff approximation.

For singlet excitations,  the excitation energies of the system are the eigenvalues of the 
  $(n_o n_c \times n_o n_c )$ Tamm-Dancoff matrix $A$ that couples excitons, i.e., occupied-virtual pairs:
\begin{equation} \label{eq:TammDan}
\begin{split}
A({i,a;j,b})=\,& (\varepsilon^{GW}_a-\varepsilon^{GW}_j)\delta_{a,b}\delta_{i,j} \\ &+2 (ia |jb )  
   - (\phi_a \phi_b |W|\phi_i \phi_j ) 
\end{split}
\end{equation}
with exchange elements
$$(ia|jb )=\int  \phi_i(r) \phi_a(r) |r-r'|^{-1} \phi_j(r') \phi_b(r') dr dr'$$
while $W \equiv W(\omega=0)$ refers to the static effective Columbic interaction approximation, and its matrix elements are
$$ (\phi_a \phi_b |W|\phi_i \phi_j )\equiv  \int  \phi_a(r) \phi_b(r) W(r,r')  \phi_i(r') \phi_j(r') dr dr'.
$$  

The superscript in $\varepsilon^{GW}$ denotes, as mentioned, that high quality \emph{GW} single particle energies should be used. In practice, and especially for medium sized systems of a few dozen electrons, it is sufficient to use the DFT eigenstates plus a GW derived correction, called a scissors approximation. We calculate the HOMO and LUMO \emph{GW} energies by the linear-scaling stochastic-GW (sGW) method \cite{Neuhauser2014sGW,Vlek20181,Vlek20182,Vlek20183, StochasticGWours,Romanova2022} and use the scissors approximation:
$  \varepsilon^{GW}_i  \simeq \varepsilon_i + \delta_o $,   
$\varepsilon^{GW}_a  \simeq \varepsilon_a + \delta_c $,
where 
$\delta_o \equiv \varepsilon^{GW}_{HOMO}-\varepsilon_{HOMO} $ 
and analogously for $\delta_c$.  Further, for higher accuracy we use the self-consistent $\Delta GW_0$ approach where the sGW HOMO and LUMO corrections are \emph{a posteriori} shifted self-consistently; for large systems this approach was found to be an excellent approximating to eigenvalue-only self-consistent $GW_0$ and to experiment.\cite{Vlek20183}  
\subsection{Mixed Representation Iterative Solution}
\vspace{-5pt}
The simplest derivation of an iterative method for  the BSE spectrum starts with the linear-response time-dependent Hartree-Fock (TD-HF) equation.\cite{Negele1982, Strinati1988,Neuhauser2005} For an initially real occupied state perturbed along the $x$ axis, $\psi_j(r,t=0) = e^{-i \alpha x} \phi_j$, one performs a linear response expansion  in $\alpha$, $ \psi_j(r,t) \simeq e^{-i\varepsilon_j t} (\phi_j(r) -i \alpha f_j(r,t))$, where $ f_j(r,t=0)=x \phi_j(r)$.  The formally non-linear TD-HF equation for $\psi_j$ then converts, for small $\alpha$, to a linear equation for $f_j$.  In the Tamm-Dancoff approximation, (where $f_j$ is not coupled to $f_j^*$) this evolution equation reads 
\begin{equation} 
i{ |\dot{f}_j\rangle} = A |f_j\rangle
\end{equation}
where 
\begin{equation} \label{eq:Aq}
    A\equiv Q \bar{A} 
\end{equation}
and
\begin{equation}\label{eq:tdse}
        \bar{A} |f_j\rangle  = 
        (H_0+ \Delta -\varepsilon_j)|f_j\rangle + (\delta v - \delta X) |\phi_j\rangle.
\end{equation}
Here we introduced several terms.  $\Delta\equiv \delta_c -\delta_o$ is the
$\Delta GW_0$ scissors shift.  The exciton Coulomb potential ia $\delta v(r,t) =  \int |r-r'|^{-1} \delta n(r',t) dr'$, where the exciton density is $ \delta n(r,t)=2\sum_i \phi_i(r) f_i(r,t)$ and the sum extends over the occupied states.  The exciton exchange $\delta X$ is defined analogously, again under the Tamm Dancoff approximation,
\begin{equation}
\langle r | \delta X(t) |\phi_j\rangle = \sum_i f_i(r,t) \int |r-r'|^{-1} \phi_i(r')\phi_j(r') dr'.
\end{equation}

Finally, $Q$ is a projection operator that ensures that the excited functions $f_j$ have no overlap with the occupied space, i.e., $Q=I-P$, with $P\equiv \sum_{s\le N_{\rm occ}} |\phi_s\rangle \langle \phi_s |$.  

The BSE equation results then when the static effective interaction $W$ replaces the Coulombic interaction in the exchange operator, yielding eventually (hiding the time-dependence of $f$):
\begin{equation} \label{eq:mixedA}
\begin{split}
     \langle r | {\bar{A}}| f_j\rangle   &  \equiv   
        (H_0 + \Delta -\varepsilon_j)f_j(r) \\ &+ \delta v(r) \phi_j(r) - \sum_i f_i(r) W_{ij}(r), 
\end{split}
\end{equation}
where the action of $W$ on the occupied-occupied term is $W_{ij}(r) \equiv \int W(r,r';\omega=0) \phi_i(r')\phi_j(r') dr'$.

The linear form of the time-dependent equation readily implies that the frequency-dependent spectrum can be obtained from the dipole-dipole correlation function, where up to a constant
\begin{equation}\label{eq:abs}
    \sigma(\omega)=\omega \bra{f^0}  \delta(A-\omega)  \ket{f^0}
    \equiv \omega \langle f^0|f(\omega)\rangle 
 \end{equation}
where $f^0_j(r)=\langle r |Q|f_j(t=0) \rangle = \langle r | Q x |\phi_j\rangle$, and the (smoothed) delta function is readily expressed using a Chebyshev expansion in $A$,
\begin{equation}
\label{eq:delta_f0}
    |f(\omega)\rangle \equiv \delta(A-\omega)\ket{f^0}\simeq\sum_n g_n(\omega) \ket{f^n}
\end{equation}
where $\ket{f^n}$ are obtained by iteratively applying a scaled $A$, starting from $\ket{f^{0}}$, while $g_n(\omega)$ are numerical coefficients.\cite{Kosloff1988,Wang1994-PRB,Weisse2006} The spectrum evaluation therefore reduces to calculation of the residues, $R_n \equiv \braket{f^0|f^n}$.  Numerically, one just requires the application of $A$ on an arbitrary exciton vector $f_j(r)$, i.e., $f\to A f$.

While here we use Eq. (\ref{eq:mixedA}), based on the mixed hole-grid representation $f_j(r)$, we note that in many cases one would want to use an explicit electron-hole basis.  For example, for systems such as large quantum dots, where $N_{\rm occ}$ is very large and we are interested in a smaller number of conduction states relative to the total number of occupied electron states, $n_c < N_{\rm occ}$, it is numerically better to replace $Q$ by a projection to the $n_c$ conduction states, $Q=\sum_{c\le n_c} |\phi_c\rangle \langle \phi_c|.$  Then, the fundamental iterated object is the electron-hole basis coefficients, $f_{ia} \equiv\langle \phi_a | f_i\rangle.$  
In the electron-hole basis, the initial state is simply the $x$-dipoles elements, $f^0_{ia} \equiv f_{ia}(t=0)= \bra{\phi_a} x \ket{\phi_i}$. 
Further, is easy to see that the iterative application of $A$ on $f$, as given in  Eqs.~(\ref{eq:Aq}),(\ref{eq:mixedA}), becomes in the electron-hole basis:
$(Af)_{i a}=\sum_{j, b} A(i,a;j,b) f_{jb}$ and $A$ is
here exactly the BSE matrix from Eq.~(\ref{eq:TammDan}).  
Practically the action by $A$ would be done then as: 
\begin{equation}
    (Af)_{ia}= (\varepsilon_a - \varepsilon_i + \Delta )f_{ia}
    +\bra{\phi_a} \delta v - \delta X \ket{\phi_j}.
\end{equation}
i.e., given the set of coefficients $f_{jb}$, one would calculate the mixed representation vectors $f_j(r)$, from which $\delta v$ and $\delta X$ follows, and then dot product per the equation above. Note that using the electron-hole basis coefficients reduces the spectral range of $A$, which will be controlled now  by highest conduction state included, instead of the (much larger) highest eigenvalue of $H_0$, thereby reducing the number of required Chebyshev terms.  Also, the same formalism carries over trivially to localized orthogonal basis sets, where $\varepsilon_a$ and $\varepsilon_i$ would be replaced by the Hamiltonian matrices within the electron and hole spaces, respectively.  The expressions for non-orthogonal basis sets can be similarly derived.

\begin{figure}
    \centering
    \includegraphics[width=2in]{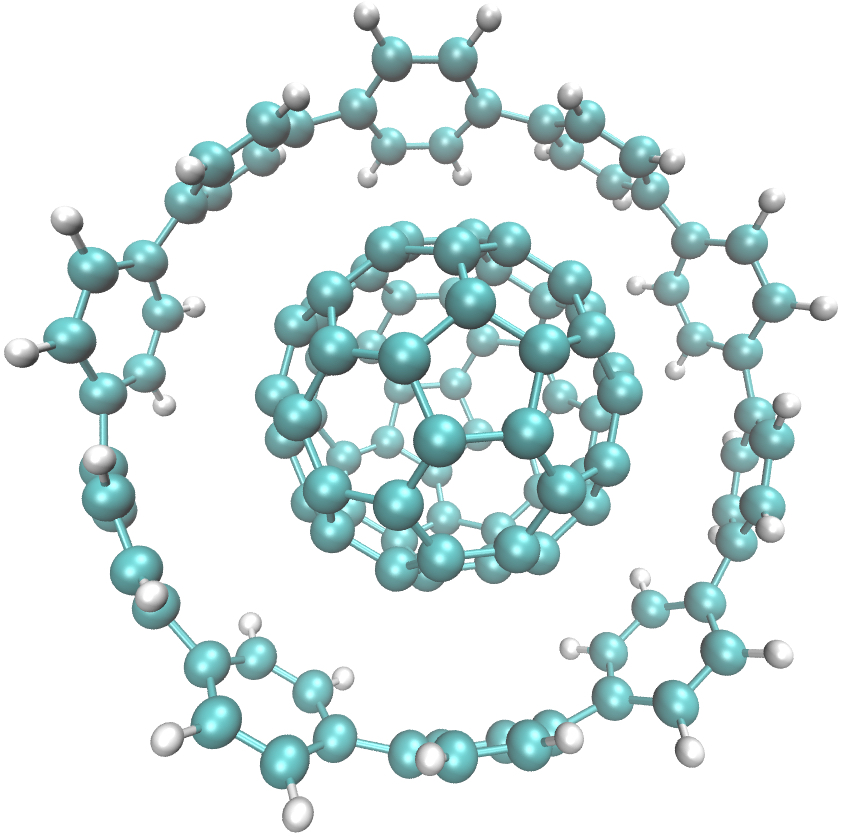}
    \caption{Structure of [10]-CPP+C$\mathrm{_{60}}$, where atomic coordinate information can be found in \cite{Minameyer2020}.}
    \label{fig:cnt}
\end{figure}

\subsection{Stochastic evaluation of the action of $W$}
\vspace{-5pt}
 The main numerical task in our formalism is the preparation of all $n_o (n_o+1)/2$ functions $W_{ij}(r)$.  $W$ is made from a static Coulomb part and a polarization component, $W=v+v \chi v\equiv v+W^{\rm pol}$ (where $v(r,r')=|r-r'|^{-1}$), so
$ W_{ij}(r)=q_{ij}+W^{\rm pol}_{ij}$, where $q_{ij} \equiv \int |r-r'|^{-1} \phi_i (r') \phi_j(r') dr'.$

As is well-known, the action of  $W^{\rm pol}$ can be obtained by time-dependent Hartree (TD-H) calculations.\cite{Louie1985,Chelikowsky2006} Specifically, for each pair of occupied functions ${1\le i,j\le n_o}$ one calculates the source ``potential" $q_{ij}(r)$ due to the $\phi_i \phi_j$ density-like source term.  Then the full set of all occupied states is perturbed, 
$\psi_s(r,t=0)=e^{-i\alpha q_{ij}(r)} \phi_s(r), s=1,...,N_{\rm occ}, $ where $\alpha \approx 10^{-6}-10^{-4} $ is a small perturbation strength, just as in the linear-response TD-HF derivation above.  Note that to avoid a plethora of indices we do not denote the dependence of $\psi_s$ on $i,j$.

The perturbed states are then numerically propagated with a TD-H Hamiltonian,
\begin{equation}
    \label{eq:dpsis_dt}
    i \dot{\psi}_s(r,t) = (H_0 + u_{ij}(r,t)) \psi_s(r,t),
\end{equation}
where $u_{ij}(r,t)$ is the potential due to the time-dependent density perturbation, 
\begin{equation}
    u_{ij}(r,t)\equiv \int |r-r'|^{-1} (n^\alpha(r',t)-n^{\alpha=0}(r',t=0)) dr',
\end{equation}
where $n^{\alpha}(r,t)=2\sum_s |\psi_s(r,t)|^2$ is the density due to the propagated perturbed orbitals.  This potential, used to propagate the time-dependent Hamiltonian, is then scaled to give the result of acting with the time-dependent effective potential, $W(t)$, i.e., 
\begin{equation} 
\bra{r} W^{\rm{\rm pol}}(t) \ket{\phi_i \phi_j} =\alpha^{-1} u_{ij}(r,t).
\end{equation} 
Finally, the desired action of the static polarization is obtained by damped integration of the action of the time-dependent polarization, $W^{\rm pol}(\omega=0)= \int_0^\infty e^{-\gamma^2 t^2/2} W^{\rm pol}(t) dt,$ i.e., 
\begin{equation} 
    \label{eq:Wijr}
    \begin{split}
    W^{\rm{pol}}_{ij}(r) = &\bra{r} W^{\rm{pol}}(\omega=0) \ket{\phi_i \phi_j} \\ &=\alpha^{-1} \int_0^\infty e^{-\gamma^2 t^2/2} u_{ij}(r,t) dt,
    \end{split}
\end{equation}
where we introduced a Gaussian damping function where the width $\gamma$ is a numerical convergence parameter.

\begin{figure}
    \centering
    \includegraphics{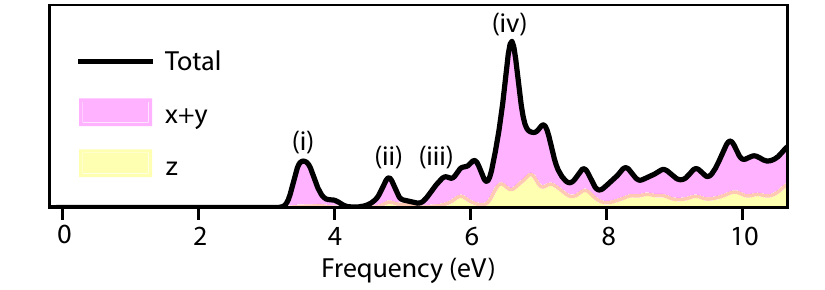}
    \caption{ Absorption spectrum of [10]-CPP+C$\mathrm{_{60}}$ calculated from an iterative solution of the BSE with a stochastic $W$.\vspace{-10pt}}
    \label{fig:abs}
\end{figure}

The one caveat in this overall approach is that the propagation of the full set of occupied orbitals is very expensive for large systems, and is in fact the most expensive portion of other large system BSE codes.\cite{BerkeleyGW,YAMBObse}  We therefore use here our stochastic-TD-H approach \cite{Neuhauser2014sGW,Gao2015} which leads to a stochastic $W$, outlined below.  (Note that this is exactly the same approach we use in our stochastic $GW$ method,\cite{Neuhauser2014sGW} with a small improvement detailed later).  Briefly, in stochastic-TD-H (or stochastic-TD-DFT in the general case) we replace the full set of occupied orbitals by a few random-sign combinations of all occupied states, 
\begin{equation}\eta_\ell (r)= {L}^{-\frac{1}{2}} \sum_{s\le N_{\rm occ} } (\pm 1) \phi_s(r),
\end{equation}
where $\ell=1,..,,L$ and for large systems a very small number of states is sufficient, $L\ll N_{\rm occ}$.  The $L$ stochastic occupied states are then treated in the TD-H procedure as if they were the full set of $N_{\rm occ}$ molecular orbitals, i.e., they are perturbed ($\eta_\ell(r,t=0^+)=e^{-i\alpha q_{ij}(r)} \eta_\ell(r)$) and propagated with $H_0+u_{ij}(r,t)$, where the time-dependent density used in constructing $u$ is now obtained from $n(r,t)=2\sum_{\ell\le L} |\eta_{\ell}(r,t)|^2$, etc.  Note that two sets need to be propagated, the perturbed $\ket{\eta^\alpha_\ell(t)} $ and unperturbed $\ket{\eta^{\alpha=0}(t)}$ stochastic orbitals.

At long times, this simple stochastic TD-H approach would eventually become unstable, due to ``contamination" by occupied states.  This means that the excited component, $\eta_\ell^{\alpha}(r,t)-\eta_\ell^{\alpha=0}(r,t)$, has in it an occupied-states' amplitudes.  For regular TD-H propagation of all states this is not a problem since in the overall density the ``contamination" of a propagated state $\psi_j(r,t)$ by an occupied $\phi_i(r)$ is exactly cancelled by the ``contamination" of the opposite pair.\cite{Zhang2015-GangTDDFT} However, in our stochastic occupied orbitals there is no such cancellation.  Luckily the instability gets tamed as the system size gets bigger, but we did find that it affects the results here if untreated for medium system sizes.  

To prevent the instability we simply ``clean" the stochastic orbitals periodically, so after every $M$'th time step we write:
\begin{equation}
    \ket{\eta^{\alpha}_\ell(t)}\to \ket{\eta^{\alpha=0}_\ell(t)} + Q\ket{\eta^\alpha_\ell(t)-\eta_\ell^{\alpha=0}(t)},
\end{equation} with $t=0, M dt,  2 M dt,... $.   This does not increase the scaling of the method since the required cleaning frequency decreases (i.e., a larger $M$ is possible) with increasing system size. Also note that after each cleaning step we renormalize each $\ket{\eta^{\alpha}_\ell(t)}$ orbital so it keeps its initial norm. In our calculations detailed below, we found $M=10$ suffices for a total time of $t=300$.

\begin{figure*}
    \centering
    \includegraphics{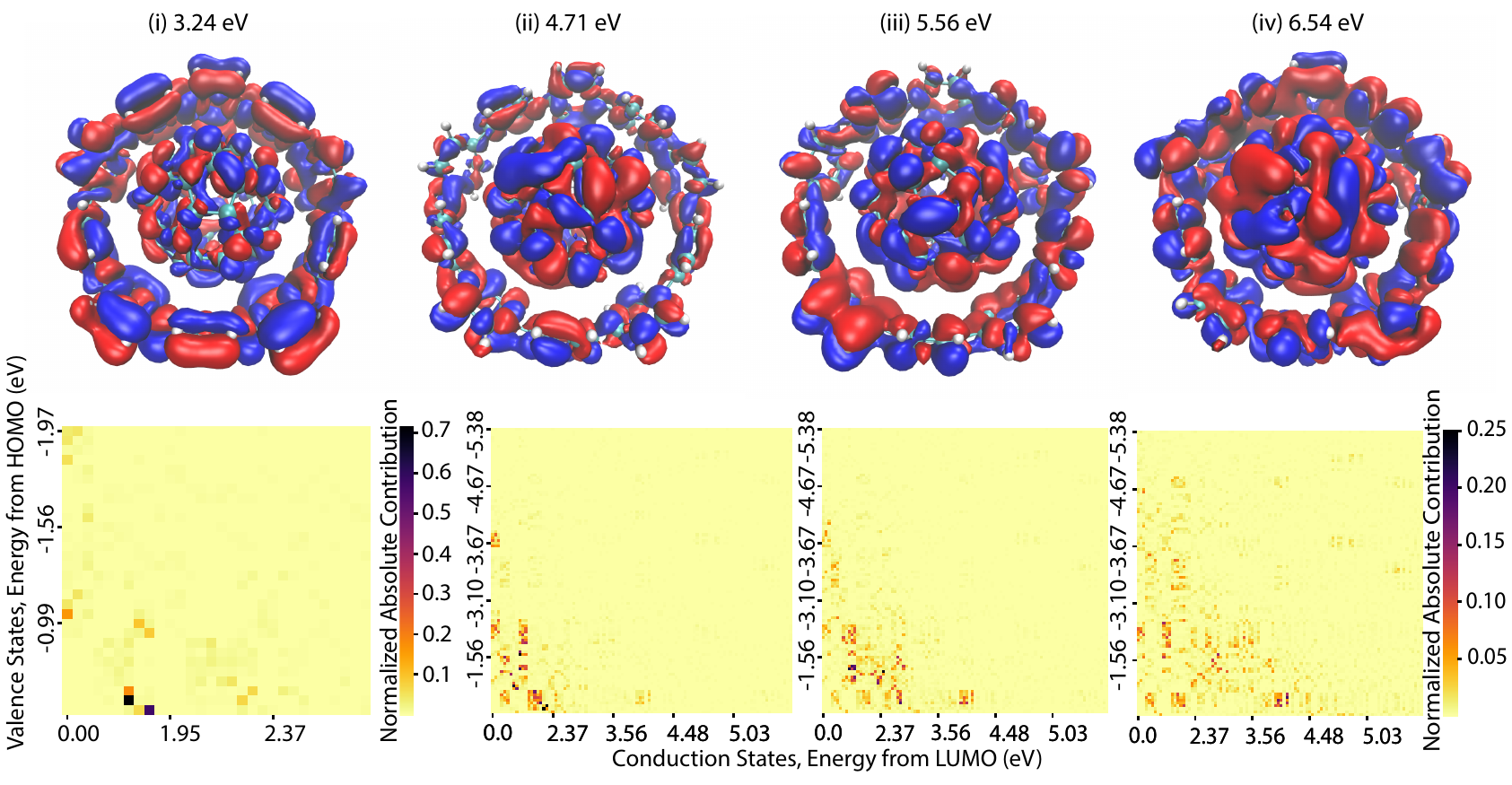}
    \caption{(Top) Exciton densities for the four lowest prominent exciton peaks, as labeled in Figure \ref{fig:abs}. (Bottom) Matrix of the corresponding (valence, conduction) transition $|f_{ia}(\omega)|$ to the exciton density above for each of these  frequencies. A small window $(n_o,n_c) = (30,30)$ is shown for the lowest exciton which includes prominent contributions from only a few electron-hole states, while a (100,100) window is used to show the higher energy excitons.  The axis are labeled by the corresponding valence/conduction energies.  The square pixels are for a given valence to conduction transition, and due to degeneracy and rising density of states, the energy labels are not linear.  Also note that while the many weaker-intensity higher $n_o$ and/or $n_c$ transitions are not visible in this colored matrix, they are important for the quantitative spectrum.}
    \label{fig:den}
\end{figure*}

Finally, a  well-known technical aspect is that due to the use of a finite grid box size, the scissors parameter needs to be shifted down by the $W(k\to0)$ term.  We use a variation of the existing procedures to find this term.\cite{Onida1995,Rozzi2006} We repeat the iterative calculation for a medium size (i.e., a lesser $n_o$ than the one eventually used) at several  different box sizes, and for each run find the average Kohn-Sham potential on the box boundary, $v^{\rm bndry}$. Then we approximate $W(k\to 0)= \epsilon_{\rm eff}^{-1} v^{\rm bndry}$, and the parameter $\epsilon_{\rm eff}^{-1}$, playing the role of an inverse dielectric constant, is fit so that the spectra from the different box-sizes runs overlap. 

Summarizing the resulting algorithm:  we use stochastic TD-H to prepare $W_{ij}(r)$ for $n_o(n_o+1)/2$ occupied-state pairs. Separately we use sGW to calculate the self-consistent \emph{GW} scissors shift $\Delta$ and subtract from it the $W(k\to0)$ term.  Then for each polarization we start with the dipole exciton state. The Tamm-Dancoff operator, Eq.~(\ref{eq:mixedA}), is successively applied and the Chebyshev residues $R_n$ are used to calculate the absorption frequencies.  As usual, if one wants to characterize the different peaks then the Chebyshev expansion of the frequency-resolved exciton state, Eq.~(\ref{eq:delta_f0}), can be used (potentially with filter-diagonalization\cite{Neuhauser1990,Mandelshtam1993} for resolving different sub-peaks).  

The two parts of the method, preparing $W_{ij}$ and applying the $A$ operator, both scale as $O(n_o^2 n_g)$ for $n_g$ grid points, more gentle than current methods.  Formally, this is  cubic scaling with system size but in practice the scaling is better since $n_o$ often rises only gently with $N_{\rm occ}$. In addition, the number of grid points would be reduced in future studies as we have shown that very sparse grids suffice when using orthogonal projected augmented waves (OPAW) instead of pseudo-potentials.\cite{Li2020}

\vspace{-10pt}
\section{Results}
\vspace{-5pt}
We demonstrate the algorithm on a characteristic large organic semiconducting system, a carbon nanohoop-fullerene complex. In the last decade, cycloparaphenylenes (CPPs), also known as carbon nanohoops,  have emerged as highly structurally tunable emitters, with rich size-dependent opto-electronic properties and host-guest chemistry.\cite{Leonhardt2019} While substantial DFT modeling has been completed on CPP+fullerene complexes,\cite{Wong2009, Yuan2015, Minameyer2020, Stasyuk2021}
extraction of optical properties at this level of theory is difficult due to the  characteristic charge transfer states in CPPs. Further, it has already been established that the BSE formalism is very accurate in predicting the properties of other fullerene-polymer complexes.\cite{Niedzialek2014} Here we present detailed results for the smallest such CPP+fullerene ``pea-pod,"\cite{Iwamoto2011, Xia2012} [10]-CPP+C$\mathrm{_{60}}$.

To simulate [10]-CPP+C$\mathrm{_{60}}$, we use a generous box of $(N_x,N_y,Nz) = (100,100,84)$ with a grid spacing of $0.5$Bohr, using norm-conserving pseudo-potentials (NCPP).\cite{Reis2003,Willand2013}  The effective inverse dielectric constant was found to be $0.30$, which gives, at this grid size, a shift of  $-W(k\to0)=-0.29{\rm eV}$.
 
 The DFT gap is 1.02eV. This gap is corrected with stochastic GW by an amount of 1.24eV, and after applying the self-consistent  self-consistent $\Delta GW_0$ this gap correction rises to 1.33 eV (i.e., a fundamental  $\Delta GW_0$  gap of 2.35eV). Combined with $-W(k\to0)$, the overall scissors shift used is $\Delta=1.04$.  

The calculations of the action of $W$, Eqs.~(\ref{eq:dpsis_dt})-(\ref{eq:Wijr}), were done with a broadening of $\gamma=0.1$ Hartree.  A time-step $dt=0.1$ a.u. was used for a split-operator propagation of $L=10$ stochastic states, and the cleaning was done every $M=10$ steps. The runs were done for $n_0=100$ valence states, requiring $n_0 (n_0+1)/2 = 5005$ actions of $W$.

The BSE iterative Chebyshev procedure was then done using the $n_o=100$ valence states.  The Chebyshev expansion of $\delta(A-\omega)$ is evaluated with a Gaussian broadening of 0.08 eV; this does not effect the spectrum significantly as it is naturally broadened due to the large number of excitons.  The runs took a total of 4000 core hours, i.e., 40 wall hours on a 100-core AMD Milan cluster.

In Figure \ref{fig:abs}, we show the calculated absorption spectrum, both total and separated to in-plane ($x$ and $y$), and perpendicular ($z$) polarizations. While there are no gas phase spectra of [10]-CPP+C$\mathrm{_{60}}$ due to the fullerene slipping out of its ``pea-pod", for the lowest exciton energy, we get reasonable agreement with just [10]-CPP,\cite{Adamska2014,Fujitsuka2012} and a stabilizing [10]-CPP+C$\mathrm{_{60}}$-[2]-Rotaxane complex.\cite{Xu2018} For just 10-[CPP] in the gas phase the lowest strong transition sits 0.4 eV higher than shown in Figure \ref{fig:abs}. Qualitatively, with the addition of the fullerene in the middle in our simulations, the overall dielectric constant would increase, thereby lowering the energy of the first exciton state. This shift is consistent with the shift to lower energies seen in the stabilized  [10]-CPP+C$\mathrm{_{60}}$-[2]-Rotaxane complex in solution.

In Figure \ref{fig:den}, we show the exciton density of four prominent excitons as labeled in Figure \ref{fig:abs}, and, for the sake of analysis we also extract the exciton density in the electron-hole state basis. Specifically, we calculate $\ket{f_j(\omega)}$ at the four exciton peak frequencies and then calculate the overlap onto to a set of unoccupied wave functions, giving us $f_{ia}$ for as many $a$ as we desire. For the lowest peak most of the exciton density is concentrated at near-gap $i,a$ states. However, for the largest spectrum peak at 6.83 eV, labeled (iv) in Figure \ref{fig:abs}, one ought to go beyond the figure and use (100,250) states to capture the same amount of density, a basis size that's very substantial. This demonstrates the strengths of the stochastic resolution of the action of $W$ in our present approach, as the full unoccupied space is sampled rather than just a subset of conduction states.

\section{Forward perspective}

Our results show that even larger systems are feasible.  This is evident by the fact that the runs were not optimised.  For example, while we used $n_0=100$, a smaller number $n_0=70$ would have sufficed, reducing the cost by a factor of two.  Similarly, the box size was very large, and a grid with almost half the points (or even less with OPAW\cite{Li2020}) would have sufficed. With optimized parameters, one can therefore easily reach systems with 1000-2000 electrons.  

An interesting feature of the method is that for larger systems it becomes less and less sensitive to stochastic errors.  Those errors appear primarily  in the sGW calculation of the scissors shift, but this calculation scales sub-linearly with system size.  Already here the sGW calculation took less than 20$\%$ of the total calculation and had an error of $0.05$eV, so for larger systems an even higher accuracy would be obtained with a small fraction of the total cost.

An interesting question is how to go to huge systems, with many thousands of electrons.  For this we note that that while this implementation of the BSE scales formally cubically, it holds promise to give eventually quadratic scaling. To achieve this, one would need to implement localized occupied basis sets sets (see references \cite{Banerjee2016,Peng2022}) for reducing the number of $i,j$ pairs for which $W_{ij}(r)$ needs to be calculated, and perhaps even stochastically sample these $\phi_i(r)\phi_j(r)$ pairs.  Work along this lines would be reported in future publications.

\vspace{-10pt}
\section*{Acknowledgments}
\vspace{-5pt}
NCB would like to acknowledge the National Science Foundation Graduate Research Fellowship Program under grant DGE-2034835. JRC would like to acknowledge the ACS Petroleum Research Fund, grant 62717-DNI6. NCB, MN, and DN are grateful for support by NSF grant CHE-1763176. Computational resources were supplied through the XSEDE allocation TG-CHE170058.

\bibliography{bse}

\end{document}